\documentclass[aps,prl,twocolumn,letterpaper,groupedaddress,showpacs]{revtex4}
\usepackage{amsmath}
\usepackage{amsfonts}
\usepackage{amssymb}
\usepackage{graphicx}
\usepackage{epsfig}

\begin{document}

\title{Breakdown of Hydrodynamics in a Simple One-Dimensional Fluid}

\author{Pablo I. Hurtado}
%\email{phurtado@onsager.ugr.es}

\affiliation{Department of Physics, Boston University, Boston, Massachusetts 02215, USA, and \\
Institute \emph{Carlos I} for Theoretical and Computational Physics, Universidad
de Granada, Granada 18071, Spain}

\today

\begin{abstract}
We investigate the behavior of a one-dimensional diatomic fluid under a shock wave excitation.
We find that the properties of the resulting shock wave are in striking contrast with those 
predicted by hydrodynamic and kinetic approaches, \emph{e.g.}, the hydrodynamic profiles relax 
algebraically toward their equilibrium values. 
Deviations from local thermodynamic equilibrium are persistent, decaying as a power law of the distance to the 
shock layer. Non-equipartition is observed infinitely far from the shock wave, and the velocity-distribution 
moments exhibit multiscaling. These results question the validity of simple hydrodynamic theories to 
understand collective behavior in 1d fluids. 
\end{abstract}

%\pacs{}

%\keywords{}

\maketitle

Most one-dimensional (1d) systems --both quantum and classical-- display anomalous
collective properties. Generally speaking, the dimensional constrain inherent 
in these systems affects strongly the propagation of fluctuations. These fluctuations, in turn, control the
system cooperative behavior, so their abnormal propagation leads to the observed anomalies.
This observation underlies most surprising results found in 1d \cite{books,reviews}.

Particularly interesting is the nonequilibrium behavior of 1d classical 
fluids \cite{reviews,SF,GH,hardpoint,Narayan,ModC,Boltz,Politi,Cercignani,piston}.
An essential feature of these systems is that the particle sequence remains unchanged during their
time evolution. This important trait, also known as \emph{single-file} constrain, is 
directly responsible for the anomalous propagation of fluctuations. 
For instance, in order for a particle in a 1d fluid to move 
appreciably, a coherent, correlated motion of many particles is needed. This strongly 
suppresses diffusion: in 1d (stochastic) fluids the mean square displacement grows now as
$\langle \Delta x^2\rangle_{\text SF} = 2Ft^{1/2}$, much slower than usual Fickian diffusion 
($\sim t$) \cite{SF}. This single-file diffusion, which turns out to be relevant in many fields ranging 
from membrane biophysics to carbon nanotubes, zeolites, DNA, polymers and nanofluids, \emph{etc.}, 
has been recently confirmed in experiments with colloids \cite{SF}. Common to all these systems is
the presence of confining structures (\emph{e.g.} narrow channels) inducing \emph{quasi}-1d (single-file) motion.

In the same way anomalous diffusion reflects the strong spatial correlations in 1d fluids, energy transport
in 1d reveals the presence of long-range velocity and current correlations.
The simplest nonequilibrium situation where this is put forward is Fourier's law. When the 1d fluid is 
subject to a small temperature gradient $\nabla T$, an energy flux $J=-\kappa \nabla T$ appears
from the hot to the cold reservoir. It has been found that the heat conductivity $\kappa$ has unusual
large values for 1d fluids. In particular, it is actually believed (though some controversy 
remains \cite{GH,hardpoint}) that any classical 1d fluid with momentum-conserving interactions and 
non-zero total pressure should exhibit a divergent $\kappa$ in the thermodynamic limit, \emph{i.e.}
$\kappa \propto L^{\alpha}$ as the system size $L$ goes to $\infty$.
Although there exist analytical and numerical evidences supporting this result in both 1d fluids and crystals, 
there is still no agreement on the theoretical framework and the exponent $\alpha>0$: fluctuating 
hydrodynamics predicts $\alpha=1/3$ \cite{Narayan}, while 
mode-coupling theories and Boltzmann equation result in $\alpha=2/5$ \cite{ModC,Boltz}, and 
numerical results are still inconclusive \cite{GH, hardpoint} (see however \cite{Politi} for new strong 
evidences supporting $\alpha=1/3$). In this way, the behavior of 1d fluids far from equilibrium and the
theoretical approach suitable for their description are still open problems.

In this Letter we study the response of a 1d model fluid to a nonequilibrium excitation.
In particular, we study the piston problem, a textbook example where we learn the notion of shock 
wave \cite{Whitham}. Our results show that the anomalous nonequilibrium response of 1d fluids cannot 
be described within a classical hydrodynamic approach. In particular, the two basic hypothesis underlying any 
hydrodynamic theory are violated in our model, namely: (i) perturbations away from local thermodynamic 
equilibrium decay algebraically, persisting for a long-time, and (ii) there are additional slowly-changing
observables apart from the density, velocity, and energy fields.

Our model is the diatomic gas of hard-point particles \cite{GH,hardpoint,Politi}.
In a line of length $L$ we introduce 
$N=L$ hard-core point particles interacting via elastic collisions. 
The particles have alternating masses, $m_{2i-1}=m=1$ and $m_{2i}=M$, $i\in [1,\frac{N}{2}]$.
In addition, we include a moving piston of \emph{infinite} mass, which starts moving from $x=0$ at
$t=0$ with constant velocity $V=1$. 
All other particles are initially at rest (cold gas)\cite{T0}, and start with random positions, provided that 
the alternating order holds. All simulations here reported correspond to systems with $N=2\times 10^4$ 
particles 
(other sizes were also simulated to check our results against finite-size effects),
and we average our results over $10^4 - 5\times 10^4$ different realizations of the initial state.

The propagation of the piston into the cold fluid creates a shock wave
traveling with velocity $W>V$. At any time $t$, the flow field consists of an aftershock region 
$Vt < x < Wt$ of excited gas attached  to the piston, and a pre-shock region $x>Wt$ 
of undisturbed gas. The position of the shock wave can be identified 
with the \emph{leading particle}, trivially defined at any $t$ as the last particle disturbed from 
its initial, motionless state. For equal masses, $m=M$, the shock structure is particularly simple. In this
case the system is equivalent to a 1d gas of non-interacting particles: the leading particle
moves with velocity $2V$, and the aftershock fluid is composed by particles with velocity $0$ and $2V$,
so the average is $V$. However, this limit is pathological, since the system is now non-ergodic,
and lacks any relaxation mechanism in velocity space capable of driving the fluid toward 
(local) equilibrium. We may restore relaxation by letting $M>m$. In this case one
expects to recover far behind the shock wave an equilibrium 
state with number density $n$, average flow velocity $V$ and temperature $T$. Conservation of
particles, total momentum and total energy, or equivalently the constancy of the corresponding
fluxes (Rankine-Hugoniot conditions), thus imply,\cite{Zeldovich}
\begin{equation}
n=2n_0, \quad W=2V, \quad T=\frac{1}{2}(m+M)V^2,
\label{asymptotic}
\end{equation}
where $n_0$ is the density of the undisturbed gas. 
\begin{figure}[t]
\centerline{
\psfig{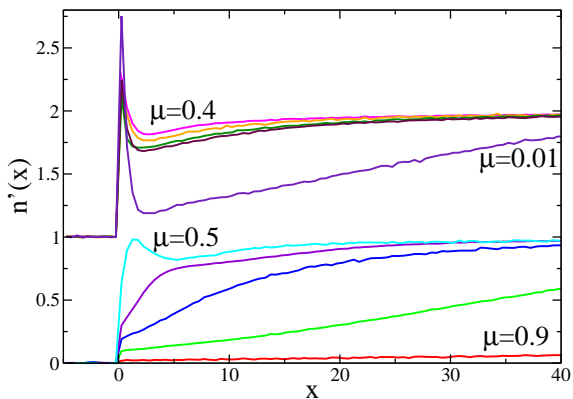}}
\caption
{\small (color online) Number density profiles for different mass ratios. From bottom to top, 
$\mu=0.9, 0.8, 0.7, 0.618, 0.5$. For $\mu<0.5$ profiles have been shifted 1 unit in the
vertical axis. From top to bottom, $\mu=0.4, 0.3, 0.2, 0.1, 0.01$.  
}
\label{profiles}
\end{figure}

The randomness of the initial positions implies that the location and velocity of the shock wave 
(equivalently, the leader) are stochastic variables when consider within the ensemble 
of initial configurations. While the average shock position grows as $\langle x_{\ell}\rangle= Wt$, 
the variance of the shock location is measured to grow diffusively, $\langle \Delta x_{\ell}^2\rangle=2Dt$, 
with $D$ an effective mobility
(see \cite{Lebowitz} for similar shock fluctuations). To prevent these
fluctuations from blurring the shock wave structure\cite{blur},  we measure the flow profiles,
characterizing the transition from the undisturbed gas toward the asymptotic equilibrium 
state (\ref{asymptotic}), in a reference frame moving with the leading particle.
Moreover, profiles quickly converge to a steady shape after a short transient, thus allowing
time averaging. Fig. \ref{profiles} shows the normalized number density profiles measured
in this way. Normalized observables are defined as $\phi'(x)=(\phi(x)-\phi_-)/(\phi_+-\phi_-)$, where
$\phi_{\pm}=\phi(x\to \pm \infty)$. 
\begin{figure}
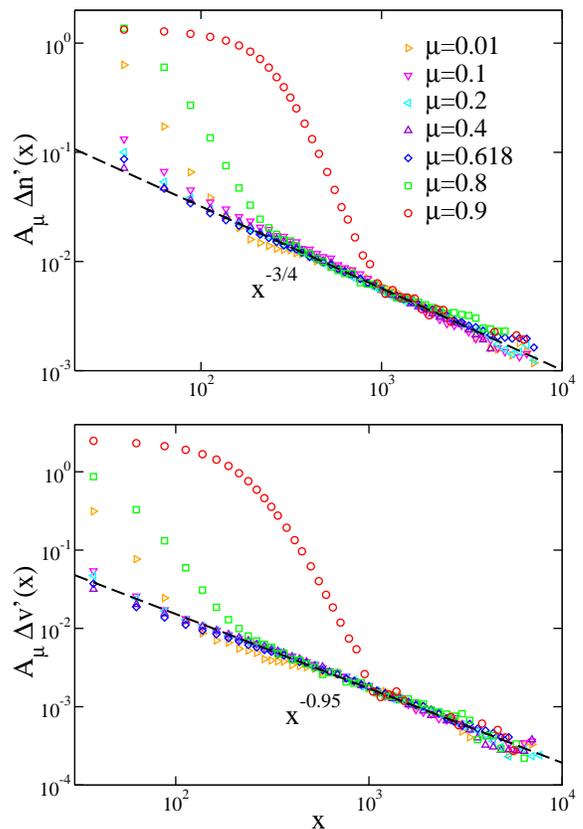

\centerline{\psfig{file=density-profiles-power-law-scaling2.eps,width=7.5cm}}
\vskip 0.3cm 
\centerline{\psfig{file=velocity-profiles-power-law-scaling4.eps,width=7.5cm}}
\caption
{\small (color online) Log-log scaling plot of $\Delta n'(x)$ and $\Delta v'(x)$.
Many different $\mu \in [0.01,0.9]$ are included on each plot. The scaling 
factors $A_{\mu}$ have been selected in each case to fix the amplitude of the power-law tail to 1. 
}
\label{dpower}
\end{figure}

Most interesting in Fig. \ref{profiles} is the dependence of profiles on mass ratio, $\mu=m/M$. 
For $\mu$ close to $1$, the profile changes smoothly; in particular, the shock width (the length 
scale on which profiles change appreciably) is much larger than the mean free path.
However, as $\mu$ decreases, the shock 
becomes steeper. This tendency lasts up to $\mu \approx 0.5$. At this point a density overshoot
develops. Further decreasing $\mu$ results in a smoothening of the profile, with the exception of the
initial overshoot, which sharpens as $\mu$ decreases. The shock width exhibits then a characteristic 
non-monotonous $\mu$-dependence, with a minimum around $\mu\approx 0.5$. Notice that 
$\mu\approx 0.5$ was previously reported to yield the fastest relaxation times for this 
model \cite{hardpoint}. Similar behavior is observed  in velocity and energy profiles (not shown). 
Interestingly enough, both density and velocity overshoots are not observed in hard-sphere
fluids for $d>1$.

For large $x$, where $x$ is the distance from the shock layer, one expects the fluid 
to relax toward the equilibrium state (\ref{asymptotic}). To characterize this relaxation,
we now measure the \emph{excess} flow fields as a function of $x$, 
defined as $\Delta \phi'(x)\equiv 1-\phi'(x)$, with $\phi'=n', v', E'$. This is shown in Fig. \ref{dpower}.
It is remarkable that these observables relax toward the asymptotic state (\ref{asymptotic}) as
a power law, $\Delta \phi'(x) \sim x^{-\beta_{\phi}}$. In particular, we find $\beta_{\text n} \approx \frac{3}{4}$, 
and $\beta_{\text v} =\beta_{\text E}  \approx 0.95\neq \beta_{\text n}$ (energy curves are not 
shown).  These algebraic tails do not depend on $\mu$, as demonstrated by the scaling in Fig. \ref{dpower}, 
although the observed transient behavior is large for $\mu$ close to 0 and 1, and minimal for $\mu \approx 0.5$.
The power laws are a reflection of the large correlations present in 1d fluids, and are in striking
contrast with the relaxation predicted by hydrodynamics and kinetic approaches, which is always
exponential \cite{Whitham, Cercignani}. In fact, this slow algebraic relaxation is related to the presence
of long-wavelength hydrodynamic fluctuations \cite{Andreev}, which control the fluid's transport 
properties. We discuss further this connection below. 

\begin{figure}[t]
\centerline{
\psfig{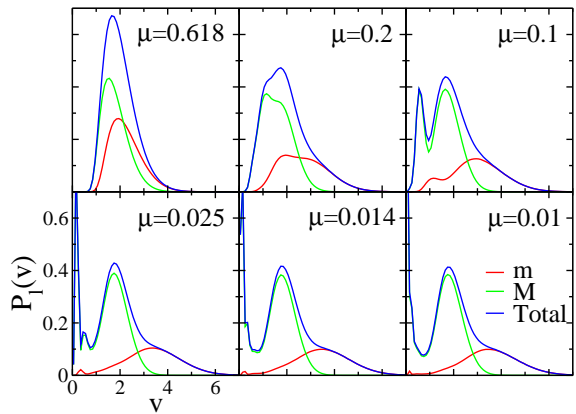}}
\caption
{\small (color online) $P_{\ell}(v)$ and $P_{\ell}^{\textrm{(k)}}(v)$, $k=m, M$ for 
different mass ratios.}
\label{distrib}
\end{figure}
We now focus on the shock wave propagation. Fig. \ref{distrib} shows the shock stationary 
velocity distribution, $P_{\ell}(v)\equiv P(x=0,v)$, for several $\mu$'s. The average 
propagation velocity is $W=2$ in all cases, as expected. However,
a multi-peaked structure in $P_{\ell}(v)$ emerges for $\mu \le 0.5$, which coincides with the 
appearance of profile overshoots, see Fig. \ref{profiles}. This exemplifies
how far the system is from local thermodynamic equilibrium near the shock layer. We may split 
$P_{\ell}(v)=P_{\ell}^{\textrm{(m)}}(v) + P_{\ell}^{\textrm{(M)}}(v)$, where $P_{\ell}^{\textrm{(k)}}(v)$ is the probability density
function (pdf) for finding a particle of mass $k=m,M$ and velocity $v$ leading the shock (or equivalently, 
the pdf $P^{\textrm{(k)}}(x=0,v)$ for having a particle of mass $k=m,M$ at $x=0$ with velocity $v$ in a reference frame 
moving with the shock wave). These distributions, plotted also in Fig. \ref{distrib}, show that the shock
wave propagates faster (slower) whenever a light (heavy) particle is leading.
Also remarkable is the incipient singularity in $P_{\ell}(v)$ for small $v$ as $\mu \to 0$. This, together with the 
singular overshoot in the profiles for  small $\mu$ (Fig. \ref{profiles}), makes the limits $\mu \to 0$  and $\mu \to 1$
essentially different, contrary to recent claims \cite{hardpoint}.
\begin{figure}
\centerline{
\psfig{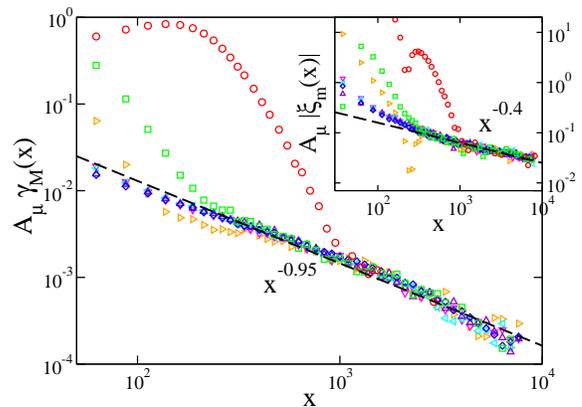}}
\caption
{\small (color online) Log-log scaling plot of $\gamma_{\text M}(x)$ (main) and $\vert \xi_{\text m}(x)\vert$ (inset) versus
the distance to the shock. Definition of amplitudes $A_{\mu}$ and values of $\mu$ as in Fig. \ref{dpower}. }
\label{kpower}
\end{figure}

\begin{figure}
\centerline{
\psfig{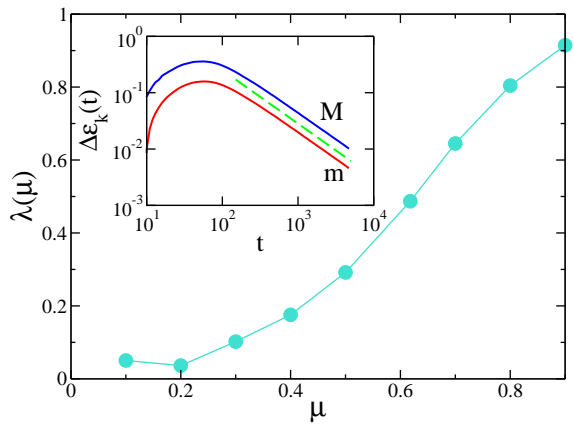}}
\caption
{\small (color online) Main: amplitude ratio $\lambda(\mu)$ (see main text for definition) as a function 
of $\mu$. Inset: Log-log plot of $\Delta \epsilon_{\text k}(t)$, $k=m, M$, vs. time for the particular case $\mu=0.8$. 
The dashed line has slope $0.95\approx\beta_{\text E}$.}
\label{etot}
\end{figure}

Next, we turn our attention to the emergence of local thermodynamic equilibrium (LTE) behind the shock. 
For that, we measure the skewness $\gamma_{\text k}(x)$ and kurtosis excess $\vert \xi_{\text k}(x)\vert$ 
of the local velocity distribution for each species, $P^{\textrm{(k)}}(x,v)$, $k=m, M$.
They are defined as $\gamma_{\text k}(x)\equiv [\nu_3^{\textrm{(k)}}(\nu_2^{\textrm{(k)}})^{-3/2}](x)$ and
$\xi_{\text k}(x)\equiv 3-[\nu_4^{\textrm{(k)}}(\nu_2^{\textrm{(k)}})^{-2}](x)$, where $\nu_i^{\textrm{(k)}}(x)$ is the $i-$th central
moment of $P^{\textrm{(k)}}(x,v)$. As the system reaches LTE behind the shock, 
both $\gamma_{\text k}(x)$ and $\vert \xi_{\text k}(x)\vert$ must go to zero. Fig. \ref{kpower} shows the actual 
relaxation. Surprisingly, we find that both quantities decay also as slow power laws, 
$\phi_{\text k}(x) \sim x^{-\beta_{\phi}}$, where now $\beta_{\gamma}\approx 0.95$ and 
$\beta_{\xi}\approx 0.4$ (exponents are independent of the species $k=m, M$).
This proves that perturbations away from LTE are persistent in 1d, and that higher-order fields other 
than the hydrodynamic ones (namely,  $\gamma_{\text k}(x)$ and $\vert \xi_{\text k}(x)\vert$) change slowly and therefore
are relevant. Also interesting is
the fact that $\beta_{\gamma} > \beta_{\xi}$. This observation, combined with the other exponents 
$\beta_{\text n,v,E}$, implies that the distributions $P^{\textrm{(k)}}(x,v)$ exhibit multiscaling.
This emerging picture is intriguing and exemplifies the anomalous behavior of 1d fluids. 
Another interesting observable is the total energy stored in light and heavy particles at a given time 
$t$, $\epsilon_{\text k}(t)$, $k=m, M$. Energy equipartition in the asymptotic equilibrium state implies that
$\epsilon_{\text k}(t) \to \frac{1}{4}(3k + \bar{k})V^2\equiv \epsilon_{\text k}(\infty)$ 
as $t\to \infty$, where $\bar{k}$ is the mass other than $k$. The inset to Fig. \ref{etot} shows the excess
energies $\Delta\epsilon_{\text k}(t) = \epsilon_{\text k}(\infty) - \epsilon_{\text k}(t)$, $k=m, M$, for a particular $\mu$. 
They also decay as a power law, as expected from the algebraic tail of $\Delta E'(x)$.
However, what is remarkable is the different amplitudes: light
particles are \emph{always} closer to the asymptotic equipartition energy than heavy particles 
(\emph{i.e.} light particles are more energetic than heavy ones \cite{GH,Cercignani,piston,Jou}). We may 
characterize this non-equipartition of energy with the ratio 
$\lambda(\mu)\equiv \Delta\epsilon_{\text m}(t)/\Delta\epsilon_{\text M}(t)$, which converges to a $\mu$-dependent
constant for long enough times. This is shown in Fig. \ref{etot}, where it is evident that energy non-equipartition
becomes more important as $\mu$ decreases. 

A full theoretical explanation of the observed behavior seems challenging. 
Both hydrodynamic \cite{Whitham} and (Boltzmann) kinetic \cite{Cercignani} approaches do not account for 
the measured shock structure and the slow, power-law relaxation of hydrodynamic observables.
These algebraic tails result from long-wavelength hydrodynamic fluctuations, a main source of non-locality 
and long-range correlations in fluids. In fact, for 3d fluids reflection and refraction of hydrodynamic fluctuations 
on a shock wave give rise to a slow, \emph{universal} power-law relaxation of \emph{all} hydrodynamic fields, 
$\Delta \phi(x)\sim x^{-3/2}$, $\phi=n,v,E$ \cite{Andreev}. Although somewhat similar in spirit, 1d fluids 
exhibit a much more intricate behavior. The algebraic tails for the hydrodynamic fields are not universal, with 
exponents $\beta_{\text n}\approx \frac{3}{4}$, and $\beta_{\text v}=\beta_{\text E}\approx 0.95\neq \beta_{\text n}$. 
In addition, we observe non-equipartition of energy and multiscaling behavior 
of one-particle distribution functions. Most importantly, deviations away from local thermodynamic equilibrium 
are persistent, as characterized by the slow power-law decay of the skewness and kurtosis fields, with exponents
$\beta_{\gamma}\approx 0.95$ and $\beta_{\xi}\approx 0.4$, respectively.  

Any hydrodynamic description of collective behavior relies on two basic hypothesis \cite{Narayan}: (i) the 
system must reach local thermodynamic equilibrium rapidly (\emph{i.e.} exponentially fast), and (ii) the only 
slowly-evolving observables are the density, velocity and energy fields. The slow algebraic relaxation of 
$\gamma_{\text k}(x)$ and $\vert \xi_{\text k}(x)\vert$ proves that both hypothesis are violated in our 1d model fluid, 
which on the other hand contains the main ingredients characterizing a broad class 
of 1d systems (\emph{i.e.} dimensional constrain, and energy and momentum conservation).
This suggests that the anomalous collective properties of 1d fluids cannot be described within a simple 
hydrodynamic approach. This is particularly important in the study of transport phenomena in these 
systems, where hydrodynamic equations have been used to predict an anomalous thermal 
conductivity, for instance \cite{Narayan}. A sound theoretical approach to collective phenomena 
in 1d fluids must therefore go beyond hydrodynamics. A good candidate is generalized 
kinetic theory. However, taking into account the strong inter-particle correlations present in 1d fluids in 
such a microscopic theory is a challenge that deserves further attention.

\begin{acknowledgments}
The author thanks P.L. Krapivsky for the original idea of this study, and T. Antal, P.L. Garrido, P.L. 
Krapivsky, and S. Redner for useful discussions. Financial support from the Spanish MEC is also acknowledged.
\end{acknowledgments}


\begin{thebibliography}{99}
\bibitem{T0} This initial state ensures an inifinitely strong shock wave, which probes the fluid's
whole nonequilibrium relaxation. A $T\ne 0$ initial state does not change our main results.
\end{thebibliography}
\end{document}